\documentclass[twocolumn,superscriptaddress,nofootinbib,preprintnumbers,showpacs,tightenlines,notitlepage]{revtex4-1}
\usepackage[latin1]{inputenc}
\usepackage{blindtext}
\usepackage{graphicx}
\usepackage{amssymb}
\usepackage{color}
\usepackage{float}
\usepackage{amsmath}
\usepackage{amsfonts}
\usepackage{dcolumn}
\usepackage{hyperref}
\usepackage{amsthm}
\usepackage{color}
\usepackage{bm}

\def\uudot{\dot{u}}
\def\3nab{\tilde{\nabla}}

\def\be {\begin{equation}}
\def\ee {\end{equation}}
\def\ba {\begin{eqnarray}}
\def\ea {\end{eqnarray}}

\newtheorem*{pro*}{Proposition}
\newtheorem*{thm*}{Theorem}

\newcommand{\bra}[1]{\left(#1\right)}

\newcommand{\brac}[1]{\left\{#1\right\}}

\newcommand{\sfr}[2]{{\textstyle\frac{#1}{#2}}}

\newcommand{\lc}{\varepsilon}
\newcommand{\E}{{\mathcal E}}

\newcommand{\barray}{\begin{array}}
\newcommand{\earray}{\end{array}}
\newcommand{\e}{e}
\newcommand{\N}{N}
\newcommand{\A}{{\cal A}}
 \newcommand{\nab}{\nabla}

\newcommand \ep {\epsilon}

\newcommand{\udot}{{\mathcal A}}
\newcommand{\hh}{{\mathcal H}}

\newcommand{\bw}{\begin{widetext}}
\newcommand{\ew}{\end{widetext}}
\newcommand{\ben}{\begin{enumerate}}
\newcommand{\een}{\end{enumerate}}

\newcommand{\bef}{\begin{frame}}
\newcommand{\eef}{\end{frame}}
\newcommand{\bi}{\begin{itemize}}
\newcommand{\ei}{\end{itemize}}

\newcommand{\bfl}{\begin{flushleft}}
\newcommand{\efl}{\end{flushleft}}

\newcommand{\bb}{\begin{block}}
\newcommand{\eb}{\end{block}}

\newcommand{\bse}{\begin{subequation}}
\newcommand{\ese}{\end{subequation}}
\newcommand{\eei}{\end{eqnarray}\indent\indent}
\newcommand{\bc}{\begin{center}}
\newcommand{\ec}{\end{center}}
\newcommand{\ber}{\begin{eqnarray}}
\newcommand{\eer}{\end{eqnarray}}
\newcommand{\bern}{\begin{eqnarray*}}
\newcommand{\eern}{\end{eqnarray*}}
\newcommand{\beast}{\begin{equation*}}
\newcommand{\eeast}{\end{equation*}}
\newcommand{\bal}{\begin{align}}
\newcommand{\eal}{\end{align}} 

\newcommand{\hs}{\,-\,}

\newcommand \om {\omega}

\newcommand{\sfrac}[2]{{\textstyle{#1\over#2}}}
\def\case#1/#2{\textstyle\frac{#1}{#2} }
\newcommand{\nb}{\nabla}

\newcommand{\bver}{\begin{verbatim}}
\newcommand{\ever}{\end{verbatim}}

\newcommand{\fR}{$f(R)$}

\begin{document}

\title{Rotating and spatially twisting Locally Rotationally Symmetric Spacetimes in \fR-Gravity: a No-Go theorem}
\author{Sayuri Singh}
\email{sayurisingh22@gmail.com}
\affiliation{Astrophysics and Cosmology Research Unit, School of Mathematics, Statistics and Computer Science, University of KwaZulu-Natal, Private Bag X54001, Durban 4000, South Africa.}
\author{Amare Abebe}
\email{amare.abbebe@gmail.com }
\affiliation{Center for Space Research \& Department of Physics, North-West University,  Mafikeng  2735, South Africa.}
 \author{Rituparno Goswami}
\email{Goswami@ukzn.ac.za}
\affiliation{Astrophysics and Cosmology Research Unit, School of Mathematics, Statistics and Computer Science, University of KwaZulu-Natal, Private Bag X54001, Durban 4000, South Africa.}

 \author{Sunil D. Maharaj}
 \email{Maharaj@ukzn.ac.za}
\affiliation{Astrophysics and Cosmology Research Unit, School of Mathematics, Statistics and Computer Science, University of KwaZulu-Natal, Private Bag X54001, Durban 4000, South Africa.}
\begin{abstract}
Recently, in a series of papers, we established the existence of and found a general solution for the simultaneously rotating and twisting locally rotationally symmetric spacetimes in General Relativity which can model inhomogeneous and dynamic astrophysical bodies. However, these spacetimes necessarily require imperfect fluids with entropy flux. Therefore, in this paper, we investigate the existence of these spacetimes in generic $f(R)$-gravity models, where the entropy flux can be generated purely by higher order curvature effects, while the standard matter still remains a perfect fluid. We transparently demonstrate here that the symmetries of these spacetimes force the theory to be General Relativity. This is a novel study that shows how the geometrical properties of a spacetime can be used to restrict the theories of gravity and how a solution can, in fact, favor a theory with fewer degrees of freedom.
 \end{abstract}
 
\maketitle
\section{Introduction}
{\em Locally Rotationally Symmetric} (LRS) spacetimes have been studied in detail in the context of General Relativity (GR) \cite{Ellis_1967,Ellis_1968,Ellis_1969,Ellis_1973,Ellis_1996}. These spacetimes possess a continuous isotropy group at each point, generally implying the existence of a multiply-transitive isometry group acting on the manifold \cite{Ellis_1967,Ellis_1968}. As a result of this isometry group, there is a preferred spacelike direction $e^a$ orthogonal to the fluid flow 4-vector $u^a$, and all spatial directions orthogonal to $e^a$ and $u^a$ are geometrically identical.

It has been rigorously shown in GR, for a perfect fluid form of matter, the LRS spacetimes cannot have rotation (of the vector $u^a$) and spatial twist (of the vector $e^a$) simultaneously \cite{Ellis_1996}. On this basis, perfect fluid LRS spacetimes can be classified into three distinct subclasses. LRS class I denotes the spacetimes that are rotating and not spatially twisting. The spacetimes in this class are necessarily stationary and inhomogeneous. LRS class II includes those spacetimes that have vanishing rotation and twist (for example spherically symmetric spacetimes). Spacetimes in this class can be both dynamic and inhomogeneous. LRS class III describes the spacetimes with vanishing rotation but non-zero spatial twist. These spacetimes are necessarily homogeneous but dynamic. 

\subsection{Simultaneously rotating and twisting LRS spacetimes}

From the above classification of perfect fluid LRS spacetimes, it is clear that none of them are suitable for describing a dynamic and inhomogeneous rotating star. Therefore we must relax the perfect fluid condition to obtain a solution for a spacetime which is both rotating and spatially twisting. Recently in a series of papers \cite{sayuri2016,sayuri2017}, it was established that for imperfect fluids (necessarily with entropic flux) it is possible to have LRS spacetimes with simultaneous rotation and spatial twist. These spacetimes exhibit novel features and also they are self similar. Using the feature of self similarity we can reduce the field equations to a set of dimensionless ordinary differential equations and constraints that can be easily solved to obtain a general solution \cite{sayuri2017}. This solution has very interesting cosmological and astrophysical properties, with the caveat that the imperfect fluids are not physically interesting.

\subsection{\fR-gravity, a possible alternative?}

One possible way to overcome the above mentioned problem of imperfect fluids is to seek a rotating and twisting LRS solution in higher-order gravitational theories like $f(R)$-gravity.
Since an imperfect fluid is a necessary ingredient for the above spacetime, one would naturally expect these to be  contained in $f(R)$-gravity for the following reason. It is well known that the field equations in these theories reduce to standard Einstein field equations with an {\it effective} two component energy-momentum tensor. The first component is generated by the standard matter in the spacetime that can be a perfect fluid, while the second one is generated by the higher order curvature terms which identically go to zero when $f(R)=R$. This so called {\it curvature fluid} is imperfect in general and hence the total effective energy-momentum tensor becomes that of an imperfect fluid even if the standard matter remains a perfect one. Therefore it is really important to investigate whether a perfect fluid along with a geometrical entropy flux in higher order gravity can produce a simultaneously rotating and twisting LRS spacetime.

\subsection{It doesn't work: a No-Go Theorem}

In the process of solving the field equations for LRS spacetimes with non-zero rotation and spatial twist in $f(R)$-gravity, we came across a very interesting result. It turns out, that the geometrical symmetries of the LRS spacetimes with simultaneous rotation and spatial twist, more specifically the self similarity, forces the underlying theory of gravity  to be GR. This no-go theorem is an interesting example where the kinematical features of a solution choose a theory with less degrees of freedom, as $f(R)$-gravity has an extra scalar degree of freedom in comparison to GR.\\

The paper is organized as follows. In the next section we discuss the semi-tetrad decomposition of LRS spacetimes. We then write down the field equations for LRS spacetimes in $f(R)$-gravity, with the effective energy-momentum tensor generated by the curvature terms. We utilize the self-similar properties of the given spacetime to reduce the field equations to a system of ordinary differential equations and constraints. Finally we show how these constraints force  the theory to be linear in the Ricci scalar.

\section{Semi-tetrad formalism for LRS spacetimes}
As a result  of the symmetries of LRS spacetimes, where there is a covariantly defined, spacelike preferred direction at each point on the manifold, the $1+1+2$ semi-tetrad covariant formalism \cite{clarkson2003,bets2004,clarkson2007} (being a natural extension of local 1+3 decomposition) is well suited for describing the geometry. In this formalism, the field equations become a set of coupled differential equations in covariantly defined scalar variables. With respect to a timelike congruence, the spacetime can be locally decomposed into time and space parts. This timelike congruence can be defined by the matter flow lines with the normalized 4-velocity $u^a$. Furthermore if the normalized vector $e^a$ is along the preferred spacelike direction, we split the spacetime metric in the following way:
\be
g_{ab}=-u_au_b+h_{ab}\equiv -u_au_b+e_ae_b+N_{ab}\;.
\ee
Here $N_{ab}$ is the metric of the 2-surfaces perpendicular to both $u^a$ and $e^a$. As by the symmetry of LRS spacetimes all the directions on this 2-surface are geometrically equivalent hence total projection of any derivatives on them must vanish. This gives rise to two covariantly defined derivatives, the \textit{covariant time derivative} along the flow lines (denoted by a dot) for any tensor 
$ S^{a..b}{}_{c..d}$, given by 
\be
\dot{S}^{a..b}{}_{c..d}{} = u^{e} \nab_{e} {S}^{a..b}{}_{c..d}\;,
\ee
and the covariant spatial derivative along the preferred direction,
\be
\hat{\psi}_{a..b}{}^{c..d} = e^{e}h^a{}_f
h^p{}_c...h^b{}_g h^q{}_d h^r{}_e \nab_{r} {S}^{f..g}{}_{p..q}\label{eq:hat}.
\ee
To obtain the required kinematical quantities of the spacetime, we first decompose the covariant derivative of the vector $u^a$ along and perpendicular to it, to get
 \be
\nabla_a u_b = -u_a A_b +\frac13\Theta h_{ab}+\sigma_{ab}+\ep_{abc}\om^{c},
\ee
where $A_b=\dot u_b$ is the acceleration, $\Theta=D_au^a$ is the expansion of $u_a$, $\sigma_{ab}=\bra{h^c{}_{\left( a \right.}h^d{}_{\left. b \right)}-\sfr13 h_{ab} h^{cd}}D_cu_d$ is the shear tensor (i.e. the rate of distortion) and $\om^{c}$ is the vorticity vector (i.e. the rotation). 
The Weyl tensor is also split relative to $u^a$ into the \textit{electric} and \textit{magnetic Weyl curvature} parts as $E_{ab} = C_{abcd}u^bu^d$ and $H_{ab} =\sfr12\ep_{ade}C^{de}{}_{bc}u^c $. Similarly, the energy-momentum tensor of matter is decomposed as 
\be
T_{ab}=\mu u_au_b+q_au_b+q_bu_a+ph_{ab}+\pi_{ab}\;,
\ee
where $\mu=T_{ab}u^au^b$ is the energy density, $q_a=h^{c}{}_aT_{cd}u^d$ is the 3-vector defining the heat flux, $p=(1/3 )h^{ab}T_{ab}$ is the isotropic pressure, and $\pi_{ab}=T_{cd}h^{c}{}_{\langle a}h^{d}{}_{b\rangle}$ is the anisotropic stress. 
We now further decompose the above quantities with respect to the preferred spacelike direction $e^a$, obtaining 
\ba
\uudot^a&=&\udot \e^a+\udot^a,\\
\omega^a&=&\Omega \e^a+\Omega^a,\\
\sigma_{ab}&=&\Sigma\bra{\e_a\e_b-\sfr{1}{2}\N_{ab}}+2\Sigma_{(a}\e_{b)}+\Sigma_{ab},\\
E_{ab}&=&{\cal E}\bra{\e_a\e_b-\sfr{1}{2}\N_{ab}}+2{\cal E}_{(a}\e_{b)}+{\cal E}_{ab},\\
H_{ab}&=&{\cal H}\bra{\e_a\e_b-\sfr{1}{2}\N_{ab}}+2{\cal H}_{(a}\e_{b)}+{\cal
H}_{ab},\\
q^a&=&Q \e^a+Q^a,\\
\pi_{ab}&=&\Pi\bra{\e_a\e_b-\sfr{1}{2}\N_{ab}}+2\Pi_{(a}\e_{b)}+\Pi_{ab}.\\
\ea
Finally, the covariant derivative of $e^a$ is decomposed in the directions orthogonal to $u^a$ into its irreducible parts, which gives
\be 
h^c{}_a h^d{}_b\nab_{c}e_{d} = e_{a}a_{b} + \frac{1}{2}\phi N_{ab} + 
\xi\epsilon_{ab} + \zeta_{ab}~.
\ee
Here, $\epsilon_{ab}=\epsilon_{[ab]}$ is the volume element on the sheet, $\phi$ is the \textit{spatial expansion of the sheet},  $\zeta_{ab}$ the \textit{spatial shear}, i.e., the distortion of the sheet, $a^{a}$ its \textit{spatial acceleration} , i.e., deviation from a geodesic), and $\xi$ is its spatial \textit{vorticity}, i.e., the ``twisting'' or rotation of the sheet. 

All the vectors and the tensors obtained after this double decomposition vanish identically due to the symmetry of LRS spacetimes. That is, we must have  \ba\label{eq:LRS}
&&\udot^a=\Omega^a =\Sigma_a={\cal E}_a={\cal H}_a=Q^a=\Pi^a=a_a=0\;,\\ 
&& \Sigma_{ab}={\cal E}_{ab}={\cal H}_{ab}=\Pi_{ab}=\zeta_{ab}=0\;,
\ea 
and all the remaining variables are covariantly defined scalars.

\section{LRS spacetimes in \fR-gravity}
We consider modified gravity models with a generalized Einstein-Hilbert action of the form \cite{buchdal1970,staro1980,de2010,capo2011,nojiri2011,capo2011,clifton2012,nojiri2017}
\be
A= \sfrac12 \int d^4x\sqrt{-g}\left[f(R)+2{\cal L}_m\right]\;,
\label{action}
\ee
where \fR~is any differentiable function of the Ricci scalar $R$, and ${\cal L}_m$ corresponds to the Lagrangian contribution of standard matter fields.
Applying the {\it variational principle} with respect to the metric $g_{ab}$, the generalized Einstein field equations read \cite{carloni2008, sot2010}
 \be\label{efes}
 G_{ab}=\frac{T^{m}_{ab}}{f'}+T^R_{ab}\equiv T^{T}_{ab} \;,
 \ee
where $T^{m}_{ab}$ is the usual energy\hs momentum tensor (EMT) of standard matter and
\be
T^R_{ab}=\frac{1}{f'}\left[\frac{1}{2}g_{ab}(f-Rf')+\nb_{b}\nb_{a}f'-g_{ab}\nb_{c}\nb^{c}f'\right],
\ee
 is the energy-momentum tensor of the so-called {\it curvature fluid}), and $T^{T}_{ab}$ is the {\it total effective} energy-momentum tensor. We use $f'\;,f''\;,f'''$ as the shorthands for the first, second and third derivatives of $f(R)$ with respect to $R$ respectively. It is worth noticing that GR is an \fR-gravity model with $f(R)=R$ as we can immediately see in that case $T^R_{ab} =0$.
Applying the $1+1+2$ semi-tetrad decomposition, as described in the previous section, on the total effective energy-momentum tensor and extracting LRS scalars we finally get \cite{nzioki2014jb}:
\ber
\mu^{T}&=& \frac{1}{f'} \left[\mu_m+\frac12 (Rf'-f)-\Theta  f''\dot{R} +f''' \hat{R}^{2}+ f'' \hat{\hat{R}}\right.\nonumber \\ 
&&\left. + \phi f''\hat{R} \right] \;,\\
p^{T}&=& \frac{1}{f'} \left[p^m+\frac12 (f-Rf') + f'''\dot{R}^{2} + f''\ddot{R}  - \A f'' \hat{R} + \right.\nonumber \\
&&\left.\frac23 \left(\Theta  f'' \dot{R} - \phi  f'' \hat{R} -  f''' \hat{R}^{2} -  f'' \hat{\hat{R}}\right) \right]\;,\\
Q^{T}&=& \frac{1}{f'} \left[Q^m-f''' \dot{R} \hat{R} + f''\left(\dot{\hat{R}}-\A \dot{R} \right)\right]\;,\\
\Pi^{T}&=& \frac{1}{f'} \left[\Pi^m+\frac13 \left( 2 f''' \hat{R}^{2} +2 f'' \hat{\hat{R}} - \phi f'' \hat{R} \right) - \right.\nonumber\\ 
&&\left. \Sigma  f'' \dot{R}\right]\;.
\eer
\section{Rotating and twisting LRS spacetimes: Field equations and solutions}

From the previous section, it is evident that the Ricci scalar $R$ acts as a dynamic scalar variable in \fR-gravity and hence the complete set of scalar variables that completely describes the system is given as
\ba\label{eq:LRSvar}
{\cal D}_1:&=&\{R, \udot,\Theta, \Omega, \Sigma, {\cal E}, {\cal H}, \mu^T, p^T, Q^T, \Pi^T,  \phi, \xi\}\,.
\ea
By the symmetry of LRS spacetimes all these scalars are functions of the curve parameters of integral curves of $u^a$ and $e^a$. Solving for these scalars will then completely determine the spacetime.

\subsection{The field equations}
 
 Decomposing the Ricci identities for $u^a$ and $e^a$ and the doubly contracted Bianchi identities gives the following field equations for LRS spacetimes.\\
\medskip \\
\textit{Evolution}:
\ba
   \dot\phi &=& \bra{\sfr23\Theta-\Sigma}\bra{\udot-\sfr12\phi}
+2\xi\Omega+Q^T\ , \label{phidot}
\\ 
\dot\xi &=& \bra{\sfr12\Sigma-\sfr13\Theta}\xi+\bra{\udot-\sfr12\phi}\Omega
\nonumber \\ && +\sfr12 \hh,  \label{xidot}
\\
\dot\Omega &=& \udot\xi+\Omega\bra{\Sigma-\sfr23\Theta}, \label{dotomega}
\\
\dot \hh &=& -3\xi\E+\bra{\sfr32\Sigma-\Theta}\hh+\Omega Q^T
\nonumber\\ && +\sfr32\xi\Pi^T\;.
\ea
\smallskip

\textit{Propagation}:
\ba
\hat\phi  &=&-\sfr12\phi^2+\bra{\sfr13\Theta+\Sigma}\bra{\sfr23\Theta-\Sigma}
    \nonumber\\&&+2\xi^2-\sfr23\bra{\mu^T+\Lambda}
    -\E -\sfr12\Pi^T,\,\label{hatphinl}
\\
\hat\xi &=&-\phi\xi+\bra{\sfr13\Theta+\Sigma}\Omega , \label{xihat}
\\
\hat\Sigma-\sfr23\hat\Theta&=&-\sfr32\phi\Sigma-2\xi\Omega-Q^T\
,\label{Sigthetahat}
 \\
  \hat\Omega &=& \bra{\udot-\phi}\Omega, \label{Omegahat}
\\
\hat\E-\sfr13\hat\mu^T+\sfr12\hat\Pi^T &=&
    -\sfr32\phi\bra{\E+\sfr12\Pi^T}+3\Omega\hh
 \nonumber\\&&   +\bra{\sfr12\Sigma-\sfr13\Theta}Q^T , \label{Ehatmupi}
\\
\hat \hh &=& -\bra{3\E+\mu^T+p^T-\sfr12\Pi^T}\Omega
\nonumber\\&&-\sfr32\phi \hh-\xi Q^T\;.
\ea
\smallskip

\textit{Propagation/evolution}:
\ba
   \hat\udot-\dot\Theta&=&-\bra{\udot+\phi}\udot+\sfr13\Theta^2
    +\sfr32\Sigma^2 \nonumber\\
    &&-2\Omega^2+\sfr12\bra{\mu^T+3p^T-2\Lambda}\ ,\label{Raychaudhuri}
\\
    \dot\mu^T+\hat Q^T&=&-\Theta\bra{\mu^T+p^T}-\bra{\phi+2\udot}Q^T \nonumber \\
&&- \sfr32\Sigma\Pi^T,\,
\\    \label{Qhat}
\dot Q^T+\hat
p^T+\hat\Pi^T &=&-\bra{\sfr32\phi+\udot}\Pi^T-\bra{\sfr43\Theta+\Sigma} Q^T\nonumber\\
    &&-\bra{\mu^T + p^T}\udot\ ,
\ea 
\ba
\dot\Sigma-\sfr23\hat\udot
&=&
\sfr13\bra{2\udot-\phi}\udot-\bra{\sfr23\Theta+\sfr12\Sigma}\Sigma\nonumber\\
        &&-\sfr23\Omega^2-\E+\sfr12\Pi^T\, ,\label{Sigthetadot}
\\  
\dot\E +\sfr12\dot\Pi^T +\sfr13\hat Q^T &=&
    \bra{\sfr32\Sigma-\Theta}\E-\sfr12\bra{\mu^T+p^T}\Sigma \nonumber \\
  && -\sfr12\bra{\sfr13\Theta+\sfr12\Sigma}\Pi^T+3\xi\hh \nonumber\\
    &&+\sfr13\bra{\sfr12\phi-2\udot}Q^T\;.
\label{edot}
\ea

\textit{Constraint:}
\be
\hh = 3\xi\Sigma-\bra{2\udot-\phi}\Omega\;. \label{H}
\ee \\

\subsection{Properties of simultaneously rotating and twisting LRS spacetimes}

Simultaneously rotating and twisting solutions to the above set of field equations, necessarily implies the condition
\be\label{omegaxi}
\Omega\xi\neq0\;.
\ee
Then we see that for these spacetimes, {\it all} scalars $\Psi$ obey the following consistency relation:
\be\label{scalarcons}
\forall \Psi, \,\,\, \dot\Psi\Omega = \hat\Psi \xi\;,
\ee
which is easily derived by noting that for any scalar $\Psi$ in a general LRS spacetime we have $\nabla_a\Psi=-\dot{\Psi}u_a+\hat\Psi e_a$ and $\epsilon^{ab}\nabla_a\nabla_b \Psi=0$.
The important point here is: the above equation (\ref{scalarcons}) applies to all scalars, and is invariant under the transformation $\tau\rightarrow a\tau,$ $\rho \rightarrow a\rho,$ where $\tau$ and $\rho$ are the curve parameters of the integral curves of $u$ and $e$ respectively. This definitely implies {\it self similarity} as the consistency relation for all the scalar variables remains invariant under simultaneous rescaling of timelike and spacelike curve parameters. 
 The above symmetries generate further constraints and hence the total set of  constraint equations  are now  $\mathcal{C}\equiv\{\mathcal{C}_1,\mathcal{C}_2,\mathcal{C}_3,\mathcal{C}_4\}$, where  
\ba
\mathcal{C}_1:= \hh& = & 3\xi\Sigma-\bra{2\udot+\frac{\Omega}{\xi}\left(\Sigma-\sfr23\Theta\right)}\Omega\,,  \label{constraint0}\\
\mathcal{C}_2:= \phi&=& -\frac{\Omega}{\xi}\left(\Sigma-\sfr23\Theta\right) \,, \label{constraint1} \\
\mathcal{C}_3:=Q^T&=&-\sfr{\sfr{\Omega}{\xi}}{1+\bra{\sfr{\Omega}{\xi}}^2}\left(\mu^T+p^T+\Pi^T\right),\label{omegaxi2} \\ 
\mathcal{C}_4:=\E &=&  \frac{\Omega}{\xi} \udot \left(\Sigma-\frac23\Theta\right)-\Sigma^2+\frac13\Theta\Sigma +\frac29\Theta^2 \nonumber \\
&&+2\left(\xi^2-\Omega^2\right)+\sfr{\bra{\sfr{\Omega}{\xi}}^2}{1+\bra{\sfr{\Omega}{\xi}}^2}\left(\mu^T+p^T+\Pi^T\right) \nonumber \\
&&-\frac12\Pi^T-\frac23\mu^T\;. \label{E}
\ea
The new constraints are derived by taking all the scalars $\Psi\in{\cal D}_1$ and using equation (\ref{scalarcons}) and the field equations. It is therefore obvious that the time derivatives of these new constraints will identically vanish using (\ref{scalarcons}) and the field equations as we are feeding the solutions back to the same system. Hence these new constraints evolve consistently in time.
 
\subsection{Solving for the LRS scalar variables}

By using the property of self similarity, we further reduce the set of independent field equations. Let us consider the set of variables
\be
{\cal D}_2 :=\brac{\udot, \Theta, \xi, \Sigma, \Omega, \phi }\subset{\cal D}_1 ,
\ee
and from the kinematical equations for LRS spacetimes
\ba
\nabla_a u_b &=&-u_ae_b\udot+\e_a\e_b\bra{\sfr13\Theta+\Sigma}+\Omega\lc_{ab} \nonumber\\&&+\N_{ab}\bra{\sfr13\Theta-\sfr12\Sigma},\\
{D}_{a}e_{b} &=& \frac{1}{2}\phi N_{ab} + \xi\epsilon_{ab},
\ea
it is clear that for any element $g\in{\cal D}_2$, we must have
\be\label{kinvar}
g(\tau , \rho) = ag(a\tau,a\rho),
\ee
as $u^a$, $e^a$, $N^{ab}$ and $\epsilon^{ab}$ are dimensionless and $\nabla_a$ has dimension of $\rho^{-1}$.
Hence without any loss of generality, all these quantities can be written as
\be\label{fform}
g \equiv \frac{g_0(z)}{\rho},
\ee
where 
\be
z = \frac{\tau}{\rho}\,,
\ee
and $g_0$ is dimensionless. 
Again, from the field equations $G_{ab}=T^T_{ab}$, we see that any member $h\in{\cal D}_3$, where
\be
{\cal D}_3 :=\brac{R, {\cal E}, {\cal H}, \mu^T, p^T, Q^T, \Pi^T }={\cal D}_1-{\cal D}_2\,,
\ee
must satisfy
\be\label{dynvar}
h(\tau, \rho) =a^2h(a\tau , a\rho).\\
\ee
Therefore, these quantities can be generally written as
\be\label{gform}
h \equiv \frac{h_0(z)}{\rho^2}.
\ee
Now the {\it dot} and {\it hat} derivatives of all the elements of ${\cal D}_1$ can be written in terms of the dimensionless variable $z$, in the following way.
Let us consider any general scalar $\Psi$ of the form 
\be
\Psi=\frac{\Psi_0(z)}{\rho^n}\,.
\ee
Then we can write 
\ber
&&\dot{\Psi}=\frac{\Psi_0(z)_{,z}}{\rho^{n+1}}\;,\\
&& \hat{\Psi}=-\frac{z\Psi_0(z)_{,z}+n\Psi_0(z)}{\rho^{n+1}}\;.
\eer
Using the equation (\ref{scalarcons}), we then derive an important result
\be\label{psieq}
\frac{\Psi_0,z}{\Psi_0}=-\frac{n\xi_0}{\Omega_0+z\xi_0}\;.
\ee
Now letting $\Psi_0=\Omega_0$ in Eq. \eqref{psieq} we get 
\be
\frac{\Omega_{0,z}}{\Omega_0}=-\frac{\xi_0}{\Omega_0+z\xi_0},  
\ee
and from $\Psi_0=\xi_0$, we get 
\be
\frac{\xi_{0,z}}{\xi_0}=-\frac{\xi_0}{\Omega_0+z\xi_0}.  
\ee
The above two equations are coupled first order ordinary differential equations for $\Omega_0$ and $\xi_0$, and the general solution is given by
\ba
\xi_0 (z) &=& -\frac{1}{Az+B}, \\
\Omega_0 (z) &=& -\frac{B}{A(Az+B)}, 
\ea
where $A$ and $B$ are constants of integration, that has to be given at any initial epoch. We must have $A\ne0$ and $B\ne0$ for the equation (\ref{omegaxi}) to be true. Using the above the general solution for equation \eqref{psieq} is now given by
\be\label{psisol}
\Psi_0(z)=\frac{C_\Psi}{(Az+B)^n}\;,
\ee
where $C_\Psi$'s are the constants of integration  that has to be supplied or determined via the constraint equations. This solution (\ref{psisol}) is the solution for all the scalars in ${\cal{D}}_1$ and hence we have the complete solution for the system. From the analysis above we come across a very important result for these spacetimes, which we state below:
\begin{pro*}
Any dimensionless scalar in a simultaneously rotating and twisting LRS spacetime must be a constant.
\end{pro*}
This proposition has far-reaching consequences on the possible theories of gravity that contains this spacetime, which will be apparent in the next section.

\section{Reconstructing the function $f(R)$: No-Go Theorem}

Having obtained the complete solution for the system, the next obvious question is, whether this is a solution for any function $f(R)$ or does this gives a further constraint on the allowed theory of gravity. To check this we use the method of reconstruction of the theory of gravity from the given solution. Let us recall from the previous section that 
\be
R=\frac{R_0(z)}{\rho^2}=\frac{C_R}{(Az+B)^2\rho^2}\,,
\ee
and from the requirement that $f(R)$ should scale like $R$, we have
\ber
&&f(R)=\frac{F_0(z)}{\rho^2}=\frac{C_{F_0}}{(Az+B)^2\rho^2}\;,\\
&&f'(R)=F_1(z)=C_{F_1}\;,\\
&&f''(R)=F_2(z)\rho^2=C_{F_2}(Az+B)^2\rho^2\;,\\
&&f'''(R)=F_3(z)\rho^4=C_{F_3}(Az+B)^4\rho^4\;.
\eer
From the fact that
\be
f'(R)=\frac{df(R)}{dR}=F_1(z)=\frac{F_{,z}}{R_{,z}}=\frac{zF_{0,z}+2F_0}{zR_{0,z}+2R_0}\;,
\ee
it becomes immediately clear that
\be
C_{F1}=\frac{C_{F_0}}{C_R}\;.
\ee
Again calculating $F_{2}(z)$, we get
\be
F_{2}(z)=\sfrac{\left(z^2F_{0,zz}+3zF_{0,z}\right)}{\left(zR_{0,z}+2R_0\right)^2}-\sfr{\left(zF_{0,z}+2F_0\right)\left(z^2R_{0,zz}+3zR_{0,z}\right)}{\left(zR_{0,z}+2R_0\right)^3} \,.
\ee
Simplifying the above, we have
\be
F_2(z)=0\implies C_{F_2}=0\;.
\ee
Similarly, it can be shown that 
\be
F_3(z)=0\implies C_{F_3}=0\;.
\ee
The above results imply that the second order and higher derivatives of the function $f(R)$ with respect to $R$ must vanish at all points in the rotating and twisting LRS spacetimes.
Stated otherwise, GR is the only theory in the class of $f(R)$-gravity theories that allows such spacetimes to exist.
Hence we have proved the following no-go theorem:
\begin{thm*}
The most general simultaneously rotating and twisting locally rotationally symmetric spacetimes can be a solution for $f(R)$-gravity iff the function $f$ is a linear function of the Ricci scalar, which implies General Relativity.  
\end{thm*}
Also by virtue of the above Proposition,
it is clear that these spacetimes are not contained in any other Scalar-Tensor theories where a dimensionless scalar field is coupled to the Ricci scalar in the gravitational action as in that case these scalars have to be constants and hence the theory has to be GR.
\section{Discussion}

Since an imperfect fluid is a necessary ingredient for the most general rotating and spatially twisting LRS spacetimes, we investigated the existence of such spacetimes in general $f(R)$-theories of gravity where the total effective energy-momentum tensor is in the form of an imperfect fluid in general. However, when we reconstructed the possible theories of gravity from the symmetries of the solution, it turned out that the only possibility is General Relativity. 

This is an interesting example, in which the geometrical symmetries of a given spacetime restricts the possible theories of gravity. Furthermore, this illustrates a more general point that a solution can in fact favor a theory with less degrees of freedom. In our case, general $f(R)$-gravity has an extra scalar degree of freedom which is not present in General Relativity.
 
 \begin{acknowledgments}
SS and RG are supported by National Research Foundation (NRF), South Africa. SDM acknowledges that this work is based on research supported by the South African Research Chair Initiative of the Department of Science and Technology and the National Research Foundation. AA  acknowledges the hospitality of the Astrophysics and Cosmology Research Unit, University of KwaZulu-Natal, during the inspiring research visit for the preparation of this work. Furthermore, AA  acknowledges that this work is based on the research supported in part by the National Research Foundation of South Africa and the Faculty Research Committee of the Faculty of Natural and Agricultural Sciences of North-West University. 

\end{acknowledgments}

\end{document}